# Fluid Velocity Slip and Temperature Jump at a Solid Surface

**Jian-Jun SHU**[*], **Ji Bin Melvin TEO** and **Weng Kong CHAN**
School of Mechanical & Aerospace Engineering, Nanyang Technological University,
50 Nanyang Avenue, Singapore 639798.

**Abstract**

A comprehensive review of current analytical models, experimental techniques, and influencing factors is carried out to highlight the current challenges in this area. The study of fluid-solid boundary conditions has been ongoing for more than a century, starting from gas-solid interfaces and progressing to that of the more complex liquid-solid case. Breakthroughs have been made on the theoretical and experimental fronts but the mechanism behind the phenomena remains a puzzle. This paper provides a review of the theoretical models, and numerical and experimental investigations that have been carried out till date. Probable mechanisms and factors that affect the interfacial discontinuity are also documented.

## 1 Introduction

The nature of the boundary condition at a fluid-solid interface has been a long-standing conundrum. Slip and temperature jump boundary conditions, representing a discontinuity in the transport variable across the interface, were first proposed close to two centuries ago in the place of the conventional 'no-slip' type boundary conditions. This is fundamentally unsurprising due to the abrupt transition in molecular structure. The modelling of gas-solid boundary conditions within a kinetic theory framework offered much insight based on molecular interactions at the surface and was supported by numerous experiments and numerical simulations. Though the possible existence of liquid slip was first reported by Helmholtz & Piotrowski [1], the appreciably smaller order of magnitude relative to transport quantities renders the effect of the interfacial jump virtually unnoticeable in large-scale liquid systems, allowing the mathematically-straightforward conventional boundary conditions to be applied without major repercussions. As micro- and nanoscale liquid systems became more commonplace, attention to the boundary condition was rekindled – studies were performed to investigate the effect it has on the overall behaviour as well as possible enhancements in device performance.

Micro- and nanoscale transport phenomena require different treatment from the macroscopic case as interactions between solid and fluid particles become more pronounced due to higher surface to volume ratio and shorter length scales. In this regime, inertial forces can typically be neglected while effects such as rarefaction, compressibility, viscous dissipation and surface energy have to be considered. For gases, the continuum model and assumption of thermodynamic equilibrium start to break down when characteristic dimensions decrease [2]. The same boundaries are less straightforward for liquid systems but the poor agreement of the conventional models with experimental findings reveals the inadequacies of these assumptions. In fact, it should be highlighted that the no-slip boundary condition originated as an assumption without any fundamental basis [3].

---

[*] Correspondence should be addressed to Jian-Jun SHU, mjjshu@ntu.edu.sg

Referring to Fig. 1, slip flow is characterised by a nonphysical quantity termed as the slip length, which is a measure of the distance beyond the surface where velocity extrapolates to zero. This provides a convenient means of quantifying slip through experiments and the study of influencing factors such as surface roughness, wetting, electrical properties, dissolved gases, and shear rates. Navier [4] proposed the following linear slip model, which relates the tangential slip velocity, $u_s$, to the shear rate at the interface

$$u_s = b \frac{\partial u}{\partial x}\bigg|_s \quad (1)$$

where $b$ denotes the slip length and $x$ is the normal from the surface pointing into the liquid. The subscript $s$ refers to the value of the variable at the surface. This basic relation is employed in experimental models to link slip to measurable macroscopic quantities.

The temperature jump condition, sketched in Fig. 1, was postulated by Poisson in the form of Eq. (2) in analogy with the slip boundary condition

$$T_f - T_w = b_T \frac{\partial T}{\partial x}\bigg|_s \quad (2)$$

where $T_w$ and $T_f$ refer to the temperatures of the wall and the gas immediately next to it, $x$ is the coordinate normal to the wall directed toward the fluid, and $b_T$ represents the temperature jump coefficient or temperature jump length.

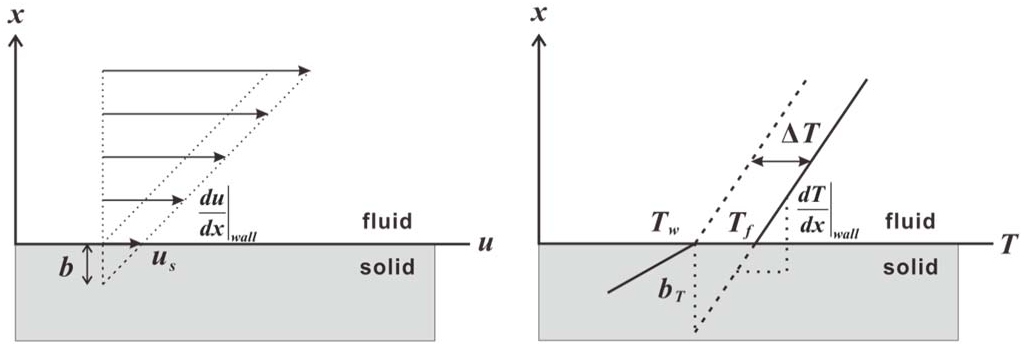

Fig. 1 Jump-type boundary conditions: (Left) slip boundary condition - $u_s$: slip velocity, $b$: slip length. (Right) temperature jump boundary condition - $\Delta T$: temperature jump, $T_w$: wall temperature, $T_f$: surface fluid temperature, and $b_T$: temperature jump length.

In microscale gas systems, the extent of deviation from the quasi-equilibrium state is measured by the Knudsen number $K_n = \frac{\lambda}{h}$, which is defined as the ratio of the molecular mean free path $\lambda$ to the characteristic domain length $h$. Typical microelectromechanical systems (MEMSs) and nanotechnology applications span the entire Knudsen regime. $K_n$ physically represents the relative dominance of molecule-wall collisions over intermolecular collisions. Slip and temperature jump effects are expected to manifest macroscopically when $K_n > 0.1$. The problem may be approached in two ways: solving the statistical Boltzmann equation or using the continuum transport equations coupled with slip or temperature jump boundary conditions. The continuum approach can provide accurate predictions in the slip regime ($10^{-3} < K_n < 0.1$). For free molecular conditions ($K_n > 10$), analytical solutions to the Boltzmann equation for simple geometries can be obtained [5] while molecular dynamics (MDs) and the direct simulation Monte Carlo (DSMC) method can provide numerical solutions for complex geometries [6]. The modelling of the transition regime ($0.1 < K_n < 10$), however, remains a problem by virtue of the equal importance of intermolecular and molecule-surface collisions. The theoretical models of gaseous slip and temperature jump are treated in the same vein – the latter is based on the exchange of momentum between the gas molecules and surface while the former considering the energy balance of the gas molecules during the scattering process.

The mechanism of Newtonian liquid slip has yet to be ascertained but two models distinguishing between true slip and apparent slip have been hypothesised [7,8]. True slip refers to the actual slipping of liquid molecules over the solid surface as opposed to apparent slip, where the sliding of liquid occurs over a less viscous layer that could be made up of a gas layer, surface coverage of nanobubbles, or even a density-depleted layer adjacent to the surface. For non-Newtonian fluids, slip has been attributed to the adhesive failure of polymer chains and disentanglement of surface chains from the bulk chains [9].

The temperature jump at a liquid-solid interface was first discovered by Kapitza [10] for superfluid helium at the temperatures of around 2K. Attempts at modelling the thermal boundary resistance using acoustic theory to describe phonon interactions at the interface have provided qualitative agreement at best. More recently, non-equilibrium MD simulations and time-domain thermoreflectance measurements have presented the evidence of temperature jump across an interface of water and self-assembled monolayer at room temperatures, revealing it to be sensitive to wetting properties, surface roughness and even the direction of heat flux. These observed dependencies are potentially useful in microscale thermal devices but are poorly understood from a theoretical perspective.

Literature on the theoretical and numerical investigations of gaseous slip and temperature jump is extensive, stemming from Maxwell's seminal work. The experimental studies of gaseous slip have largely been confined to flow rate measurements through microconduits, which contain the deleterious sources of uncertainties in the measurement of channel height and flow rate. Most liquid slip length measurement techniques are unsuitable for gas flows owing to the low magnitudes of measurable quantities while velocity mapping for gas flows is comparatively less well established. The viability of the atomic-force microscopy (AFM) as a technique for gaseous slip measurements has recently been explored [11,12].

At present, there is no direct technique that is capable of measuring liquid slip velocity or slip length. Popular experimental measurement techniques include the drainage force and tracer imaging methods. The drainage force method can be used with either the surface force apparatus or atomic force microscope, which possess high resolutions but at the same time are susceptible to experimental artefacts such as cantilever stiffness [13] and contamination. Velocity tracking methods have comparatively poorer resolution. The lack of a benchmark has seemingly led to conflicting results being reported [14]. The slip length measurement uncertainties of 2nm in drainage force methods have been claimed [15]; this is still somewhat unsatisfactory for smaller slip lengths such as that of water on mica which is roughly 20nm. There is room for improvement in the areas of resolution and reliability of slip length measurements before any empirical work on boundary slip can be deemed conclusive.

The theory of fluid-solid boundary conditions is currently lacking as most models are incapable of predicting experimentally observed results. The major drawbacks of the present models include the use of phenomenological constants and the use of separate models for gas-solid and liquid-solid interfaces. Some lingering questions that remain unanswered include the nonlinear shear rate-dependent slip, influence of wetting, near-wall molecular structure, and dependence on surface temperature. These shortcomings serve as motivation for this paper, where we aim to develop unified analytical models that are capable of describing the boundary jump phenomena for both gas and liquid systems [16,17].

## 2  Significance

The interfacial boundary condition is not only fundamentally important but also increasingly relevant in a wide range of fields, where it is of paramount interest in modern applications involving MEMS, microfluidic devices, biological systems, and colloidal chemistry.

Fluid slip plays a crucial role in myriad applications. One archetypal advantage of slippage is the reduction of flow resistance in microchannels, which is also associated with the increase in the permeability of porous media. The efficiency and pump head of microscale viscous pumps, used in drug delivery systems and microelectronic cooling, vary with the degree of slip [18,19]. The consideration of slip is important in hard disk drives as the gaseous flow at the slider head-disk interface typically lies in the slip and transitional regime. Due to its nanoscale order of magnitude, fluid slip possibly has unrealised potential applications especially in nanochannels.

The temperature jump finds uses in heat transfer applications like microcooling for electronic devices, micro heat exchangers, and fuel cells. In thermal management applications, a low thermal boundary resistance is desirable for increasing heat dissipation in microelectronic cooling while a high resistance could act as a thermal barrier. Large temperature jumps may even have potential novel uses in temperature shielding and as a form of passive temperature control. The recent discovery of the thermal rectification effect shows promise for the development of fluid-based thermal logic components [20].

The majority of the latest studies in this area have been concentrated on investigating the effects of wetting and surface roughness, specifically with the use of superhydrophobic surfaces which are artificially patterned to allow the pockets of dissolved gases and also chemically coated to reduce wettability [21,22]. Such surfaces have the ability to induce high slip velocities and temperature jumps arising from secondary slip processes. A key issue that remains elusive is the true physical mechanism of the boundary jump. This involves the consideration of factors such as molecular interactions, lattice configuration of the substrate, and near-wall molecular structure of the fluid. Coupled with the maturing of atomic manipulation techniques, tunable slip and temperature jump on designer lattices may be realised in the near future [23,24].

## 3  Mechanism of Fluid Slip on Solid Surfaces

The physical process of slip remains vague despite the plethora of experimental and theoretical studies. A fairly clear picture of gas-solid slip can be derived within the kinetic theory framework. In liquid-solid slip however, the scattering model is inadequate as the situation is confounded by the intertwining of additional interactions with liquid molecules from the bulk flow. At this stage, the contentious influences of surface nanobubbles and wetting in experiments, among other factors, have to be isolated before the primary mechanism(s) can be identified. Nevertheless, several plausible slip models have been put forward.

*Scattering mechanism*

In the billiard ball model of collisions between fluid and solid molecules, the nature of reflections governs the efficiency of the net momentum exchanged during the impacts. Maxwell conjectured that the transfer of tangential momentum occurring, during diffuse but not in specular reflections, preserved the original velocity. The notion of diffuse reflections is somewhat fuzzy, but may be the thought of as the fluid molecule undergoing several collisions with the solid molecules before escaping at the same velocity as the solid. Defining the slip velocity as the mean velocity of near-wall particles (usually within a layer thickness of one mean free path), a higher proportion of specular reflections results in higher slip velocity. This description is appropriate under rarefied conditions as fluid-solid collisions are prevalent in the vicinity of the surface due to the longer mean free paths. For the denser fluids with shorter mean free paths, the contribution of scattering to slip is expected to be less dominant as fluid-fluid interactions become more important. The Maxwell model also fails to consider inelastic interactions that are intermediate between specular and diffuse reflections.

*Surface slip*

Another model of slip depicts the actual motion of liquid molecules on the bed of solid molecules. This perspective is related to the induced structural ordering of near-wall fluid molecules. Adsorbed fluid molecules that are pinned in the wells of the substrate potential induce the rearrangement of neighbouring fluid molecules due to short-range interactions, forming epitaxial layers next to the surface [25]. The regular structure is expected to be more significant in crystalline surfaces due to their periodic potential. A solidlike phase of water molecules on a mica surface has been observed experimentally using X-ray reflectivity, revealing density oscillations spanning a few monolayers [26].

On a continuum scale, slip can be visualized as the interfacial fluid layer being dragged along the boundary by adjacent layers under shear. In fact, the evolution of slip should begin at the fluid-fluid interface where the bulk fluid and top-most epitaxial layer meet with the ordered layers beneath being initially locked [27]. With increasing shear, the layers start to cleave gradually in a top-down sequence, culminating in the slip of the bottom-most fluid layer past the surface. This represents a macroscale interpretation of slip. Zooming in further to the molecular details at the interface, the slipping of the interfacial layer can be pictured as the surface diffusion with a net drift, comprising a series of hops by the fluid molecules between substrate lattice sites while being subjected to an external field [28].

There is some scepticism about the molecular slip model, as it has been estimated that a very high shear rate of about $10^{12} \, s^{-1}$ is needed [8]. However, adopting the rate theory model where the hopping occurs by thermal vibration shows that it is not necessary for the hydrodynamic force to be greater than the dispersion forces for the slip to occur.

A further issue has been brought up with regard to the interpretation of surface molecular motion as a continuum slip condition [29]. Distinguishing between conditions at a boundary and boundary conditions, it was advocated that the correct slipping plane congruent with a continuum assumption should be at the edge of the boundary layer where mean molecular motion converges to a bulk effect.

*Apparent slip*

Certain microscale phenomena such as the electrical double layer in electrokinetics that exhibit the large velocity gradients within a thin boundary layer may also be represented using an apparent slip velocity [30]. This approach simplifies the hydrodynamic analysis by allowing the use of continuum governing equations along with effective boundary conditions that account for the mesoscopic slip effect across the interfacial layer. The presence of a less viscous layer sandwiched between the surface and bulk flow has also been suggested as a possible cause of the anomalously high slip lengths observed in experiments [31]. Possible film types may constitute dissolved gases, coating of nanobubbles, or a density-depletion layer, the last of which has recently been disputed [32]. Though the above forms of apparent slip do not arise from the true motion of liquid molecules relative to the surface, they may be exploited as the artificial approaches of inducing low interfacial friction.

*Non-Newtonian slip*

The slip of non-Newtonian fluids, in particular polymer flows, may be explained using polymer dynamics, which also provides a viable analog for the experimentally observed shear rate dependence of Newtonian fluids [33]. Entanglement states between moving bulk flow polymer chains and surface-grafted chains give rise to three primary slip regimes [9]. In the no-slip regime of low shear rates, the bulk polymers remain locked to the surface polymers. At the critical shear rate, bound polymers begin to detach from the stretched surface polymers, resulting in the relative sliding of bulk and surface layers. The sliding velocity in this regime increases with increasing shear rate. Upon complete disentanglement, slip reaches its maximum and remains constant thereafter since the bulk flow has effectively disassociated from the surface polymer layer.

## 4 Factors Affecting Slip

The primary mechanism that drives slip may be unresolved but factors displaying an ostensible effect on the measured slip length have been identified through experiments and numerical simulations. Among the most investigated factors is the unusually large slip length of superhydrophobic surfaces which possess high contact angles owing to the combination of patterned roughness and surfactant coating. Another controversial factor is the influence of shear rate, particularly the nonlinear change in slip lengths, which could open up more avenues to potential applications. Isolating any individual factor in experimental studies is a challenging task since some of them might actually be complementary or even originate from the identical physics of molecular interactions.

*Surface roughness*

Contrary to intuition, roughness does not always act to reduce slip velocities. Richardson [34] was one of the first to suggest that roughness suppressed slippage and the macroscopic no-slip boundary condition originated from surface roughness. Dussan & Davis [35] showed that the assumption of a no-slip boundary condition resulted in a stress singularity at the moving contact line for a two-phase fluid flow. Experimental work on micro-/nanostructured surfaces has produced inconclusive results. While negative slip lengths have been measured on grooved surfaces [36], the claims of corrugation-induced drag reduction have also been reported [37-39]. Using the lattice Boltzmann (LB) simulations, Sbragaglia *et al*. [40] was able to capture the concerted effect of roughness and hydrophobicity on drag reduction. MD simulations by Ziarani & Mohamad [41] showed that slip velocity decreased monotonically with increasing roughness and there was no significant change in slip behaviour between the different topographic shapes of roughness. However Cottin-Bizonne *et al*. [42] found that nanometre-scale roughness resulted in reduced friction and were able to formulate a simple expression for the effective slip length of alternating strips of different slip lengths.

Vinogradova & Yakubov [43] tried to address these discrepancies through their own experimental findings, whereby they concluded that the confusion over how roughness influences slip may have arose from the different definitions of the wall location, whether on the bottom or top of the corrugations, prescribed by researchers. They demonstrated that using a correction factor for the wall location restored the no-slip condition for rough surfaces. On the other hand, experimental studies on carbon nanotube coated surfaces revealed that the slip length increased with increasing roughness length scale in the Cassie state but remained constant with minimal slip in the Wenzel state. (Wenzel state refers to a wetting phase where liquid penetrates the roughness cavities while liquid in the Cassie state sits above the air pockets. Transition between both states occurs at a critical contact angle $\theta_c$; liquid penetrates and fills the voids to minimise surface energy.) The validity of the Cassie and Wenzel theories has been questioned although it is generally agreed to be applicable under specific conditions [44-49]. The results of a recent study on corrugated hydrophobic surfaces have demonstrated transient slip behaviour, changing from partial slip to no slip after a few hours [50]. This coincides with the transition from a Cassie to Wenzel state as observed from the direct visualization of trapped air pockets. Current experimental efforts have been concentrated on biomimetic-inspired superhydrophobic surfaces due to the enhanced slip observed on these artificially structured surfaces [51-54].

The effect of surface roughness on slip is hard to quantify in theory since it involves not only the competition between multiple length scales but also the local flow conditions. Analytical models have been derived for the macroscopic roughness of a periodic [55-59] and random nature [60,61]. Atomic-scale corrugations, however, necessitate the consideration of the influence of dispersion forces on the near-wall arrangement of the fluid molecules. Following the fluctuation-dissipation theorem approach of Barrat & Bocquet [27] to obtain the interfacial friction coefficient, Priezjev & Troian [62] was able to demonstrate good agreement for molecular-scale corrugations between their model, which considered both fluid-solid intermolecular interaction and increased potential energy arising from roughness, and MD simulations.

*Wetting*

The initial hypothesis that slippage would only occur on the surfaces of low wettability due to the perceived weaker fluid-solid attraction was refuted after several experimental studies showed that slip was also present on completely wetted surfaces [63]. Ho *et al.* [64] presented MD simulation results showing the evidence of slip at a wetting boundary and furthermore demonstrated that the slip velocity could even increase with decreasing hydrophilic contact angle. It appeared that the equilibrium site separation played a part in allowing slip to occur. A water molecule had a higher tendency of migrating to a nearer neighbouring equilibrium site, resulting in larger slip.

The Blake-Tolstoi theory predicted the qualitative trend of higher slip velocities with increasing contact angles due to the superior mobilities of liquid molecules on a nonwetting surface [65,66]. Voronov *et al.* [67] carried out a dimensional analysis based on data from MD simulations and realised that different fluid-solid pairs did not share the same slip lengths despite having similar contact angles, as was evident in their earlier work [68]. Their results illustrated that slip lengths may not always increase with a greater contact angle and that the relative molecular sizes of the fluid and solid should also be considered.

*Near-wall fluid molecular structure*

Early computational work on the epitaxial layering of near-wall fluid molecules has led to the investigation of its relationship with slip [69]. It should be noted that the near-wall ordering is indirectly linked to molecular-scale roughness, in terms of the potential exerted on the fluid molecules, and wetting, which can be ascribed to fluid-solid affinity. The dependence of slip on molecular structure is not straightforward, given the counter-intuitive ability of a solidlike phase to produce stronger slip when fluid-solid molecular interaction is weak [27]. Fluid monolayers experience weaker frictional forces compared to the molecules belonging to the bulk phase with the freedom to manoeuvre themselves, causing intermolecular jamming or locking to the substrate [70]. Slip can be envisaged to occur *via* a shear melting mechanism of the monolayers that is typically observed in confined fluids, beginning from the outermost fluid-fluid layers and eventually propagating to the fluid-solid interface [71,72].

On nonwetting interfaces, a depletion layer of lower local density is thought to be a contributing factor toward apparent slip [73]. Assuming that the viscosity remains constant, the apparent slip length can be estimated by

$$b = \int_{liquid} \left[ \frac{\rho_{bulk}}{\rho_s(z)} - 1 \right] dz \qquad (3)$$

where $\rho_{bulk}$ and $\rho_s(z)$ refer to the liquid densities in the bulk flow and depletion layer.

The thin depletion layer between 5 and 20Å measured for hexadecane gives a slip length of approximately 5Å, which fails to account for the large values of up to 350nm observed for similar interfaces in experiments [32,74]. Hence, the assumption of depletion-enhanced slip may only hold for strongly hydrophobic surfaces.

*Dissolved gases*

Slippage on hydrophobic surfaces has been associated with a thin layer of low viscosity fluid or vapour lying on the surface [75-77]. Andrienko *et al.* [78] suggested that the fluid undergoes a prewetting transition during flow, generating a macroscopically thick gas film at the wall due to phase separation. The discovery of nanobubbles forming on hydrophobic surfaces from direct AFM measurements has lent credibility to this idea [79-81]. However, the effective slip in the case of isolated nanobubbles is expected to be smaller than that for a gas layer since the boundary flow is thought to alternate between the regions of complete slip (over the nanobubbles) and no-slip (over the surface). There have been suggestions that nanobubbles could also be responsible for shear-dependent slip [31]. Interestingly, nanobubbles have been detected for water on mica, which is a completely wetting interface [82]. This might offer an alternative explanation for observed slip on hydrophilic surfaces. Some authors have previously reported lower slippage for degassed liquids, which inhibit the growth of such bubbles [83,84]. On the other hand, some studies suggested that the meniscus shape of nanobubbles played an important role and that it was possible for a surface covered with bulgier nanobubbles to exhibit a no-slip boundary condition [85-88]. Though nanobubbles are generally undesirable in experimental slip measurements, they may allow for the possibility of controllable apparent slip since the fractional coverage of nanobubbles can be varied by temperature and solvent concentration.

The simple analytical two-phase models by de Gennes [31] and Tretheway & Meinhart [73], which considered the presence of a surface gas layer, estimated the slip lengths of about 7μm. This result has two ramifications: (i) it may help to explain the atypically large slip lengths observed in certain experiments [54,89] (ii) the potential to induce the enhanced slippage with a low viscosity surface film. It was shown that a fractional surface coverage of nanobubbles of around 40% is sufficient to generate slip lengths lying in the micrometre range.

*Shear rate*

Another puzzle that remains to be solved is the dependence of slip behaviour on shear rate. This phenomenon was first discovered by Thompson & Troian [90] in their MD simulations of the Couette flow of a Newtonian liquid. At low shear rates, the results were consistent with the linear Navier slip boundary condition. Beyond a certain shear rate, the slip length began to increase nonlinearly with shear thinning being ruled out as a possible cause. Thus, the assumption of a constant slip length in experimental models may not apply to the nonlinear regime. The shear rates under consideration in MD simulations are generally too high to be realised experimentally. Nonetheless, several researchers have reported the evidence of shear-dependent slip, while others have maintained that their results obey the linear Navier expression [91].

AFM measurements by Craig *et al*. [92] exhibited an obvious variation of the slip length with the approach velocity (proportional to surface shear rate) of a colloidal probe toward a planar surface. Furthermore, the no-slip behaviour at the low driving rates of the probe offers a plausible reason for the absence of slip flow in previous experiments. Zhu & Granick [93] obtained similar results at shear rates below the onset of shear thinning and even the observed large slip lengths of up to 2μm. It should be emphasised that the validity of the constant slip length model as applied in the above experiments is questionable.

It was subsequently suggested by de Gennes [31] that the reduced hydrodynamic drainage forces leading to interpretations as shear-dependent slip were possibly due to the shear-induced nucleation of nanobubbles on the surfaces. At high shear rates, the nanobubbles may be compressed into a thin film carpeting the solid surface, over which the liquid slips.

Prior to this, Spikes & Granick [94] had also proposed their own drainage force slip model pertaining to an assumption that boundary slip manifested only upon exceeding a critical shear stress value and the ensuing slip length remained constant. Empirical fits revealed that these critical values are typically small and thus may not have been detected in earlier studies. However, their model did not demonstrate an adequate fit at higher shear stress values. The nanobubble mattress model developed by Lauga & Brenner [95] agreed fairly well with the experimental data of Zhu & Granick [93] but was based on the assumption of 99% surface coverage of nanobubbles.

While the MD simulations of Thompson & Troian [90] yielded unbounded slip behaviour, Martini *et al*. [96] found that asymptotically limiting slip could be obtained by changing the wall model from one with fixed wall atoms to another that allowed for thermal motion. Hyväluoma & Harting [86] also observed decreasing slip with increasing shear rate in their LB simulations, showing that highly deformed bubbles did not produce greater slip. Gao & Feng [97] proposed that this was due to the pinning of bubbles on the edge and showed that increasing slip with shear rate could still result otherwise, depending on the flow conditions. The experimental results of Ulmanella & Ho [98] from nanochannel flow measurements too hinted at a limiting value of slip velocity at high shear rates.

## 5 Modelling of Gas Slip

The modelling of gaseous flow in the slip regimes encompasses both intermolecular interactions between gas molecules in the form of governing equations and gas-surface molecular interactions in the form of boundary conditions. Continuum governing equations coupled with appropriate slip conditions are convenient for theoretical analysis but not sufficiently robust to describe slip flow at high $K_n$ due to nonequilibrium effects. In such cases, the statistical Boltzmann equation is able to describe the ballistic fluid behaviour. The prevailing slip models used for gases are the Maxwell-type collision models. Another recent interpretation of the fluid-solid interaction involves the use of gas adsorption concepts.

*Maxwell slip model*

The slippage of gases occurs when the minimum characteristic length scale is comparable to the mean free path of intermolecular collisions ($K_n > 10^{-3}$). In this regime, wall-molecule collisions dictate the gas flow while intermolecular collisions are almost negligible. Maxwell [99] proposed that the impact of gas molecules on a wall produced two kinds of collisions – specular and diffuse. In a specular reflection, the tangential momentum of the fluid molecule is conserved while during a diffuse reflection, the equilibrated fluid molecule is re-emitted with a tangential momentum equal to that of the wall. By convention, the fraction of molecules undergoing diffuse reflections is represented by the tangential momentum accommodation coefficient (TMAC) $\sigma$, and that of specular reflections by $1-\sigma$.

The mean tangential momentum flux $p''$ at the outer boundary of the Knudsen layer of one mean free path thickness is given by the sum of the incident flux $p_i''$ and reflected fluxes $p_{sp}''$ and $p_{diff}''$

$$p'' = p_i'' + (1-\sigma) p_{sp}'' + \sigma p_{diff}'' \qquad (4)$$

where the specular flux $p_{sp}'' = -p_i''$ and diffuse flux $p_{diff}'' = -p_w'' = 0$.

Eq. (4) is reduced to

$$p'' = \sigma p_i''. \tag{5}$$

The momentum flux can be evaluated from the following expression:

$$p'' = \iiint v' p(c') f_s(c') dc' \tag{6}$$

where $f_s(c')$ refers to the velocity distribution function and $c'$ denotes the velocity vectors $u'$, $v'$, and $w'$.

By using suitable approximations for the velocity distribution functions, the slip velocity can be obtained as

$$u_s = \frac{2-\sigma}{\sigma} \alpha \frac{du}{dy} + \beta \frac{dT}{dx} \tag{7}$$

where the first term on the right represents the slip due to the velocity gradient normal to the surface and the second term is that due to the temperature gradient along the surface, also known as thermal creep.

Alternatively, a less rigorous derivation can be achieved based on the mean tangential velocities of surface gas molecules. The postcollisional tangential velocity $u_r$ can be defined as

$$u_r = \sigma u_w + (1-\sigma) u_i \tag{8}$$

where $u_w$ and $u_i$ are the average tangential wall and incident velocities, respectively.

At the wall, half the molecules can be assumed to be reflected while the other half make up the incident population. The average tangential velocity of the gas molecules at the wall $u_{av}$ is hence given by the expression

$$u_{av} = \frac{u_r + u_i}{2} = \frac{\sigma}{2} u_w + \frac{2-\sigma}{2} u_i \tag{9}$$

Considering that each molecule traverses one mean free path $\lambda$ between each collision on average, $u_i$ may be expressed as a Taylor expansion of $u_{av}$ [100]

$$u_i = u_{av} + \lambda \left. \frac{\partial u}{\partial y} \right|_{wall} + O(\lambda^2) \tag{10}$$

where $y$ is the coordinate normal to the wall.

Finally, the slip velocity $u_s$, which is defined as the difference between $u_{av}$ and $u_w$, is obtained as

$$u_s = \frac{2-\sigma}{\sigma} \lambda \left. \frac{\partial u}{\partial y} \right|_{wall} \tag{11}$$

or in a nondimensional form

$$U_s = \frac{2-\sigma}{\sigma} K_n \left. \frac{\partial U}{\partial Y} \right|_{Y=0}. \tag{12}$$

The variations of the above method such as using higher order expansions and $\frac{2}{3}\lambda$ instead of $\lambda$ in Eq. (10) have been proposed to improve the accuracy of the model at moderately large $K_n$ numbers. Nevertheless, this continuum approach is not applicable for the transition and free-molecular regimes. It is also noted that the above slip velocity expression contains a singularity in the absence of diffuse reflections, which hypothetically occurs on an atomically smooth surface. Other criticisms of the Maxwell formulation include the neglect of inelastic scattering and assumption of a constant TMAC value instead of a local value that should be determined by conditions at the location of impact.

*Langmuir slip model*

An alternative slip model based on the gas-solid interactions as described by the Langmuir's theory of adsorption of gases on solids has also been proposed [101,102]. One fundamental difference between the Maxwell and Langmuir model lies in the treatment of the wall. In the former, the wall is assumed to be a macroscopic flat surface while the latter considers discrete sites that each interact with a single atom. A TMAC-like parameter $s$ accounts for the fraction of incident interacting gas molecules that are adsorbed and subsequently desorbed at the same velocity as the wall. Correspondingly, the fraction of specular-type interactions is given by $1-s$. The mean velocity of surface molecules is

$$u_{slip} = (1-s)u_g + s\,u_w. \tag{13}$$

$s$ can be obtained from adsorption isotherms such as that of Langmuir

$$s = \frac{\beta p}{1+\beta p} \tag{14}$$

where $p$ is the hydrostatic pressure, $\beta = \dfrac{k}{k_B T}$ with $k$ being a function of the gas-solid interaction parameters.

The resultant expression for the dimensionless slip velocity is

$$u_{slip} = \frac{1}{1+\bar{\beta}p} \tag{15}$$

where $\bar{\beta} = \dfrac{\omega K_n}{4}$, $\omega$ is a function of the equilibrium constant, the local temperature, and the heat of adsorption.

Myong [102] extended the model to consider the dissociative adsorption of diatomic gas molecules which required two adjacent vacant sites and thus had a second-order dependence on the surface coverage. The Langmuir model exhibited slightly improved agreement with experimental results for nitrogen gas flows as compared to the Maxwell model but it is not mentioned if the dissociative adsorption of the nitrogen molecule actually occurs on the surface used in the experiment.

The use of adsorption concepts in boundary slip provides physical meaning to the Maxwell's phenomenological accommodation coefficient. However, the assumption of pure scattering and adsorption events using the ideal Langmuir isotherm does not present significantly new ideas with regard to slip behaviour. Extensions to the adsorption model allow the representation of effects such as nonlinear behaviour that has been observed in experiments.

## 6 Modelling of Liquid Slip

Current liquid slip models can be broadly categorised into apparent and molecular slip models. The apparent slip models attempt to provide phenomenological resolution for anomalous empirical findings that do not fit the Navier slip model while the molecular theories describe slip behaviour using the finer physical details of molecular interactions that take place at the interface.

*Two-phase model*

The uncharacteristically large slip length that were obtained in experiments, particularly on nonwetting surfaces, led to conjectures that it was due to a less viscous layer sandwiched between the liquid and surface. For a low-viscosity film of thickness $\delta$ on the solid surface, the slip length $b$ can be estimated from the matching of shear stress at the fluid-fluid interface as [55]

$$b = \delta \left( \frac{\eta}{\eta_s} - 1 \right) \qquad (16)$$

where $\eta$ and $\eta_s$ are the bulk and film viscosities.

Alternatively, for a gas layer, de Gennes [31] considered a kinetic theory expression of the shear stress of a gas and obtained an approximate slip length of

$$b = \frac{\eta}{\rho v_n} \qquad (17)$$

with $\rho$ and $v_n$ denoting the gas density and thermal velocity component normal to the surface.

Taking into consideration, the possibility of slip occurring both at the gas-solid and gas-liquid interfaces, Tretheway & Meinhart [73] worked out the apparent slip length by applying the gas slip velocity boundary conditions at the two interfaces, showing that the slip velocity was greatly enhanced under such circumstances. However, parameters such as the surface coverage of nanobubbles for intermittent coverage and film thickness are hard to quantify.

*Blake-Tolstoi model*

Tolstoi [103] was among the earliest to adopt a molecular kinetics approach for describing slip behaviour by considering the difference between surface and bulk liquid molecular mobilities. A major contribution of the model was to show a link between slip and surface wettability. His work was later improved by Blake [65] to overcome its limitations in complete-wetting situations. The Blake-Tolstoi slip length expression reads

$$b = \sigma \left\{ \exp\left[ \frac{\alpha A \gamma_{LV} (1 - \cos\theta)}{k_B T} \right] - 1 \right\} \qquad (18)$$

where $\sigma$ is the centre-to-centre molecular separation, $\alpha$ is the fraction of the surface occupied by solid, $A$ is the effective molecular surface area, $k_B$ is the Boltzmann constant, and $T$ is the temperature.

The Blake-Tolstoi model provided an adequate qualitative prediction of slip behaviour [66]. Two debatable aspects of the theory are the use of a macroscopic form of the activation energy for the molecular mobility and validity of considering a velocity gradient across a one-molecule thick layer. Other shortcomings of the theory include the difficulty in the estimation of the surface fraction parameter and the neglect of surface roughness.

*Surface diffusion model*

Ruckenstein & Rajora's [75] work was often quoted in the literature for their insightful suggestion that a surface gas layer could be a contributing factor toward the experimentally observed magnitudes of slip that otherwise could not be purely explained by their surface diffusion model. Yet, the attempt to associate the slip with the thermally activated motion of molecules on a substrate lattice deserves more plaudits. Lichter *et al.* [104] suggested a similar surface hopping mechanism in their rate theory model of slip flow. The Arrhenius-type model was conceptually similar to the previously introduced Blake-Tolstoi model but considered tilted potential barriers between the adsorption sites of the substrate with the barriers being lower in the direction of the external field. This leads to a net drift velocity, which can be considered to be the molecular slip velocity

$$U_{\text{slip}} = v_0 \lambda \exp\left(-\frac{E_0}{k_B T}\right) \sinh\left(\frac{\Delta E_{\text{shear}}}{k_B T}\right) \qquad (19)$$

where $v_0$ is the jump rate of each adsorbed molecule, $E_0$ is the potential energy barrier, and $\Delta E_{\text{shear}}$ is the change in the potential energy barrier due to shear stress exerted on the adsorbed molecules.

Though the slip velocity in Eq. (19) appears to show a nonlinear dependence on slip, a rough estimate using appropriate parameters revealed that slip remained within the linear regime for the range of experimental shear rates [105]; the expression recovers the familiar Navier form when $\Delta E_{\text{shear}} \ll k_B T$. Slight adaptations to the model have also been made to include a critical shear stress criterion and shear-dependent dissipation at high shear rates to improve the match with results from numerical simulations but lack strong physical justifications [106-108].

*Variable-density Frenkel-Kontorova model*

The dynamics of liquid molecules at solid surfaces may be modelled classically as a stochastic process using the Langevin equation for a single-molecule description [90]. More exact models such as the Fokker-Planck equation include the use of probability density functions of stochastic variables but have to be solved using numerical means in most cases.

The one-dimensional Frenkel-Kontorova (FK) model has been used to represent the molecular mechanism of slip arising from the interplay of liquid-liquid and liquid-solid interactions [109,110]. A modified form of the FK equation was proposed to account for the mass flux in the direction normal to the surface, where the near-wall density is higher due to molecular ordering. Their variable-density Frenkel-Kontorova (vdFK) equation reads

$$m\ddot{x}_i = -\frac{2\pi g}{\lambda}\sin\left(\frac{2\pi \dot{x}_i}{\lambda}\right) + k\left(\dot{x}_{i+1} - 2\dot{x}_i + \dot{x}_{i-1}\right) + \eta_{LL}(V - \dot{x}_i) - \eta_{LS}\dot{x}_i \qquad (20)$$

where the subscript $i$ is the molecular index, $m$ is the molecular mass, $g$ is the strength of the periodic potential of the substrate, $V$ is the speed of the adjacent fluid layer, $\eta_{LL}$ and $\eta_{LS}$ are the liquid-liquid and liquid-solid friction coefficients. The second term on the right-hand side represents the stochastic movement of molecules between the surface and adjacent fluid layers.

The vdFK equation qualitatively predicts the overall transition of slip: (i) no slip to the local slip with isolated molecules performing individual hops to adjacent sites (ii) local to global slip where the fluid molecules forming the surface layer move in tandem (iii) limiting slip at high driving forces.

Realistically, the relative solid-liquid and liquid-liquid affinities are expected to be dissimilar; so, the stochastic term should instead be represented as an asymmetric random walk with a net drift in the direction of the stronger attractive force.

In summary, the theoretical models described above are only the individually capable of predicting slip behaviour under specific circumstances. Ultimately, the aim for theoreticians would be to develop a model that encompasses various determining factors that have been discovered in experiments. In sections 7 and 8, we list some of these popular slip measurement techniques.

## 7 Measurement of Liquid Slip

The advent of high resolution measurement systems has boosted the precision that is required in experimental slip flow studies. This has fostered progress in the understanding of liquid slip where extremely fine measurements are involved. Yet, discrepancies may arise due to the highly sensitive nature. Experimental slip lengths have also been found to be generally larger than numerical predictions. The possible sources of experimental errors include the presence of dissolved gases and electrokinetic effects. None of the current techniques can strictly be classified as direct methods – the closest being velocity tracking methods. Instead, the slip variables are inferred from macroscale quantities such as hydrodynamic forces and flow rates. The comprehensive reviews of the present techniques are available in the literature [7,8,22,111,112]. Here, various techniques are briefly described and assessed.

*Surface force methods*
The popular surface force methods, using either surface force apparatus or atomic force microscope, transpired from the Vinogradova's model for the thin film lubrication with the consideration of slip on the two approaching surfaces [55]. The drainage force methods possess high resolutions, limiting slip length uncertainties to within 2nm. Additionally, a wide variety of surfaces are possible with AFM. Nevertheless, the technique of surface force apparatus is known to be susceptible to contamination while the AFM measurements complicated by certain factors like roughness and inertial effects [7].

The credibility of AFM measurements has been put into question due to the inconsistency in the slip lengths of polar liquids on smooth hydrophilic surfaces that were measured by various researchers [38,92,113]. Henry & Craig [114] revisited their earlier experiments in an attempt to investigate the discrepancy and discovered that the shape of the cantilever could have been the cause. The rectangular cantilevers used by Honig & Ducker [113] were superior in the repeatability of measurements to the v-shaped cantilevers used by Craig *et al*. [92], thus masking the no-slip boundary condition in the latter case. A separate study by Rodrigues *et al*. [13] identified other experimental factors such as cantilever stiffness, approach velocity, and liquid viscosity.

*Tracers*
The most straightforward way to measure slip is through the flow visualisation with the aid of tracer particles. Such studies have been performed using microparticle imaging velocimetry, total internal reflection velocimetry, total internal reflection using fluorescence recovery after photobleaching (TIRF), fluorescence cross-correlations, and thermal motion of tracers. The imaging techniques usually suffer from low resolution due to uncertainties in determining the wall and particle positions. Moreover, the accuracy may be hampered by electrophoresis and electrostatic interactions [115]. Coupling TIRF with a Langevin-based correction method, Li *et al*. [116] recently achieved a significantly improved precision of 5 nm that is almost comparable to that of the surface force methods.

An interesting method based on the theoretical model of the relationship between the bulk diffusivity of tracers and slip velocity by Lauga & Squires [47] was employed by Joly *et al*. [117]. This technique eliminates the need for a flow source and so avoids the influence of gas bubbles. Furthermore, the fact that slip was indirectly observable in the experiments appears to rule out shear rate dependence, although it should be pointed that a more appropriate term for the effect of shear rate from a molecular framework is an external force which, for instance, can arise in the presence of a chemical potential gradient.

*Flow rate measurement*

Slip lengths may be evaluated by measuring either the mass flow rate or differential pressure across a micro- or nanochannel [53,63,98], akin to that widely used in the measurement of gaseous slip. Though the experiments are relatively simple to carry out, the method suffers from low resolution, which may not be adequate for slip lengths on the order of nanometres. Besides, the extraction of slip length becomes more complex with the consideration of surface roughness and wetting properties.

*Other methods*

Besides the above methods, slip has also been examined using quartz crystal oscillators [118-121], particle sedimentation [83], increase in potential difference across a capillary containing an electrolyte solution [122]. Rheological techniques also offered a convenient means of testing with the use of viscometers [39,54,123,124] although the unusually large slip lengths of a few hundred micrometres and actual experimental uncertainty as reported by Choi & Kim [54] were doubted [125].

The contrasting slip lengths obtained for similar liquid-surface interfaces using different measurement techniques highlights the work cut out for experimentalists in this field. In fact, inconsistencies exist even within the same technique. Furthermore, current experimental uncertainties are still too large to be able to categorically distinguish between slip and no-slip behaviour. The search remains for a robust and accurate method – achievable by improving the resolution and sieving out interferences in the current methods or devising a new technique altogether.

*Numerical methods*

Numerical simulations offer a means of circumventing the complexities and challenges involved in conducting the benchtop investigations of slip. The mesoscopic LB simulation, based on the discretisation of the Boltzmann equation on a lattice, has been used for studying slip through the simulation of microflows. Though the simulations do not provide molecular-scale insight, the coarser time and length scales are closer to that of experimental conditions and therefore can be understood from a more familiar macroscale perspective. The LB simulation, however, requires an *a priori* slip generating mechanism through artificial parameters that account for boundary scattering probabilities, fluid viscosity, and interfacial properties [126].

To probe the fundamental physics at a fluid-solid boundary, MD simulation is the de facto computational tool employed in slip studies for the classical treatment of flow that is based on the Newton's equations of motions for a molecular ensemble. A potential model, such as the modified Lennard-Jones potential, determines the intermolecular interaction. This allows the variation of liquid-solid interaction strength and densities so that factors like the wettability can be controlled. In addition, the effect of near-wall molecular structures can also be observed from the simulations. The evidence of induced epitaxial layering extending a few molecular layers from the wall was found and increased structuring led to smaller slip lengths [27,69].

Despite the present-day accessibility to powerful computational resources, MD simulations face restrictions in terms of particle numbers and are only capable of dealing with length and time-scales on the nanoscale order. System conditions, for instance, the extremely high shear rates in a Couette flow setup, can neither be replicated in experiments for validation nor translated to the more useful continuum regime. Besides, inherent ambiguities with regard to the specifications of interaction potentials, wall models and thermostatting controls have been shown to affect the trend of slip behaviour [96,127,128].

## 8 Measurement of Gaseous Slip

The main experimental technique for the determination of the gaseous slip coefficient (or TMAC) is flow rate measurement under controlled low-pressure conditions [129], from which the degree of slip can be determined by choosing an appropriate value of TMAC to fit the analytical flow rate curves to the measurements. The minute mass flow rates are sensitive to small variations in temperature and surface corrugations. Slip is also alternatively quantified in the literature by the Poiseuille number $fR_e$, where $f$ is the Fanning friction factor and $R_e$ is the Reynolds number [130].

The spinning rotor gauge, originally developed for vacuum pressure measurements, has also been used for determining the TMAC from the relationship between the torque and angular velocity of the levitated sphere [131]. Again, these measurements are highly sensitive to surface conditions and temperature, which could lead to disagreements between experimental and theoretical results [132].

The surface force technique that is widely employed in liquid slip length experiments has recently been adopted for investigating the slip behaviour of air confined between glass surfaces [11]. Since the drag forces are much lower for gases, high sensitivity had to be ensured by selecting a cantilever of low stiffness and high quality factor. It is worth noting that the reported uncertainty was higher than those of the well-established mass flow rate measurements [133]. Nevertheless, this versatile technique is attractive because of its excellent controllability and furthermore avoids the meticulous process of microchannel fabrication.

The DSMC method developed by Bird [134] is a popular computational tool for simulating rarefied gas flows. In this Boltzmann equation-based stochastic approach, molecular motion and collisions are decoupled using an algorithm that samples collisions at every time step to recalculate the new molecular velocities. A caveat of this method is that the accuracy depends greatly on the collision models being employed. Generally, the DSMC method will be most effective in gas flows with $K_n$ values falling near the onset of the transitional regime, where the continuum-based methods are no longer suitable but not in highly rarefied flows where MD simulations will be more appropriate [135].

## 9 Mechanism of Temperature Jump

The imperfect energy accommodation of gas-solid interactions leading to a temperature jump is analogous to that in gaseous slip. The liquid-solid temperature jump is however thought to be due to the transport of heat carriers known as phonons across the interface.

*Molecular scattering mechanism*

The scattering model of temperature jump is similar to the kinetic theory-based model of boundary slip flow except that it considers the incomplete exchange of energy between fluid and wall molecules during collisions [136]. Again, temperature jump *via* this mechanism is expected to dominate in gases as a consequence of the longer mean free paths.

*Phonon transmission mechanism*

The existence of a boundary thermal resistance or equivalently a temperature discontinuity has been put down to the interfacial transport of phonons, which are the main carriers of thermal energy arising from the collective vibrations of atoms or molecules in nonmetals. Sound typically travels at a velocity that is an order of magnitude higher in solids compared to liquids. Going from one medium to another, the abrupt change in molecular structures as represented by the mismatch in acoustic properties creates a large impedance that prevent incident phonons from propagating freely across the interface [137-139]. This disruption of the transmission of energy is reflected as a temperature jump. In spite of the qualitative agreement, theoretically predicted thermal resistances are typically much larger than that observed in experiments, which hint that other mechanisms may be at work.

## 10 Factors Affecting Temperature Jump

The same factors that influence slip have also been found to affect the temperature jump behaviour. With the use of MD simulations, the magnitude of the temperature jump shows a dependence on the surface roughness and wettability of the surface.

*Surface roughness*

An enhancement in thermal conductance with an increase in nanoscale roughness has been found in nonequilibrium MD simulations. This has been intuitively attributed to the larger solid-liquid contact area, evident from the amplified thermal conductance for a sinusoidal roughness geometry compared to grooved corrugations and the smaller temperature jumps for taller nanopillars [140,141]. The problem is compounded by the inclusion of the effect of roughness on wetting characteristics. A larger temperature drop was observed at a rough surface which is hydrophobic but at a smooth surface when it is hydrophilic [142].

*Wetting*

The temperature jump at hydrophobic interfaces has been shown to be larger than that at hydrophilic interfaces [140-143]. A smaller thermal resistance is commonly associated with the strong hydrogen bonding between water and surfactant molecules. Shenogina *et al.* [144] obtained a simple relationship showing that the thermal conductance was proportional to $1+\cos\theta$ with $\theta$ being the contact angle. Near a hydrophilic surface, the ordered layer of liquid molecules is thought to minimise the mismatch in structure, hence allowing for the more efficient transmission of phonons [145]. In addition, Xue *et al.* [146] identified an exponential dependence on the solid-liquid bond strength for hydrophobic surfaces, whereas hydrophilic surfaces displayed a power law dependence. As the temperature jump in the nonwetting situation is consistently two to three times larger than in wetting situations across several experiments, it was suggested that the disparity could be ascribed to a less dense liquid layer analogous to that in apparent slip flow [143].

*Direction of heat transfer*

Interestingly, the thermal conductance has been discovered to be higher when heat flows from the solid to liquid phase and lower in the opposite direction. A possible reason for this phenomenon is the strong temperature dependence of the hydrogen bonds between water molecules that cause a drop in hydrogen bonds as temperature increases [147] although the MD results of Shenogina *et al.* [144] showed the augmented rectification with stronger wetting for the same surface temperature. Murad & Puri [148] demonstrated that thermal rectification could be controlled by the near-wall liquid molecular structure through either modifying wetting properties or applying an external field. The diodelike behaviour could be promising for nanoscale thermal applications.

## 11  Modelling of Gas-Solid Temperature Jump

Smoluchowski [149] developed the earliest theory of temperature jump, drawing inspiration from the Maxwell's slip theory. The thermal accommodation coefficient $\sigma_t$ represents the fraction of reflected or re-emitted molecules possessing the mean energy of gas molecules at the same temperature as the wall [136]. It can be expressed as

$$E_i - E_r = \sigma_t (E_i - E_w) \tag{21}$$

where for the $\Gamma_m$ grams of incident gas molecules crossing a unit area per second, $E_i$ refers to the total energy of incident molecules, $E_r$ is the energy of reflected and re-emitted molecules, and $E_w$ is the energy of gas molecules if they were emitted at the wall temperature. An accommodation coefficient of one may be interpreted as a molecule undergoing repeated collisions with the wall and finally getting re-emitted as if it were from a gas at the wall temperature. In contrast, a molecule that is reflected immediately on impact can be thought of as having an accommodation coefficient of zero. In effect, the accommodation coefficient merely categorises molecules into those that fully equilibrate to the energy of the wall and those that retain their original energy. Temperatures may be used in the place of energy although this is not strictly true for polyatomic gases due to their additional internal degrees-of-freedom.

The terms in bracket on the right of Eq. (21) are given by

$$E_i - E_w = \frac{k}{2} \frac{\partial T}{\partial n} + \frac{c_v p(\gamma+1)}{2\sqrt{2\pi RT}} (T_0 - T_w) \tag{22}$$

where $k$ is the thermal conductivity of the gas, $c_v$ is the specific heat at constant volume, $p$ is the pressure of the gas at the wall, $\gamma$ is the ratio of specific heats, and $R$ is the specific gas constant. The first term on the right denotes the energy possessed by the incident gas molecules for thermal conduction while the second represents the difference in translational and internal energy carried by gas streams at temperatures $T_0$ and $T_w$.

The left-hand side of Eq. (21) represents the energy transferred to the surface and is equivalent to the overall heat conducted by the gas as follows:

$$E_i - E_r = k\frac{\partial T}{\partial n}. \qquad (23)$$

Substituting Eqs. (22) and (23) in Eq. (21) and rearranging, the temperature jump can be expressed as

$$T_0 - T_w = \frac{k(2-\sigma_t)\sqrt{2\pi RT}}{\sigma_t c_v p(\gamma+1)}\frac{\partial T}{\partial n} = \frac{2\gamma\lambda(2-\sigma_t)}{\sigma_t P_r(\gamma+1)}\frac{\partial T}{\partial n} \qquad (24)$$

where the Prandtl number $P_r$ and mean free path $\lambda$ have been introduced.

The accommodation coefficients for the translational and rotational energies have been reported to be much larger than that for the vibrational energy [150]. The above derivation for polyatomic gas molecules does not distinguish between accommodation coefficients for translational, rotational, and vibrational energies. A more rigorous approach would be to consider Eq. (21) for each energy component as in the anisotropic scattering model of Dadzie & Méolans [151].

A classical calculation of the accommodation coefficient by Baule [152] using the conservation of linear momentum and energy for the $n$ number of elastic collisions between a monoatomic gas molecule and surface of respective masses $m_g$ and $m_w$ gives the following expression:

$$\sigma_t = 1 - \left[\frac{m_g^2 + m_w^2}{(m_g + m_w)^2}\right]^n. \qquad (25)$$

According to Eq. (25), the accommodation coefficient decreases when the mass of one is much larger than the other. A higher accommodation coefficient occurs for a rough surface, on which a gas molecule may impinge repeatedly before being re-emitted.

For more highly rarefied gases, higher order temperature gradient terms are expected to exert greater influence on temperature jump. Deissler [153] derived a second-order form of the temperature jump boundary condition, additionally taking into consideration the distributions of molecular velocity and angles of incidence of the impinging gas molecules. It was also proposed that a distinction be made between the mean free path for translational energy exchange and that for internal energy exchange. The fully developed two-dimensional second-order temperature jump expression is as follows:

$$T_0 - T_w = \frac{2\gamma\lambda(2-\sigma_t)}{\sigma_t P_r(\gamma+1)}\frac{\partial T}{\partial n} - \frac{9\lambda^2(177\gamma-145)}{128(\gamma+1)}\frac{\partial^2 T}{\partial n^2}. \qquad (26)$$

For laminar heat transfer in cylindrical tubes, the first-order and second-order solutions only diverge at $K_n = 0.1$, differing by approximately 15% when $K_n = 0.2$. The second-order terms account for the nonlinear constitutive relation between heat flux and temperature gradient when the mean free path is on the order of the characteristic length. In this state, both intermolecular and molecule-wall collisions have to be considered so that the correct solution can only be obtained using molecular-based models. For instance, a closed-form solution of the linearised Boltzmann equation was obtained by Lees & Liu [154] for the heat transfer of a monoatomic gas between parallel plates.

**12 Modelling of Liquid-Solid Temperature Jump**

The two main models of the Kapitza resistance considered phonon interactions at an interface between dissimilar media. In the acoustic mismatch model (AMM) [155], the low-temperature phonon transmission probability is a function of the contrasting acoustic impedances while it depends on the balance of phonon density of states in the diffuse mismatch model (DMM) for the thermal resistance at solid-solid interfaces [138]. The different derivations originate from the assumptions of fully specular reflections in the AMM and diffuse reflections in the DMM. Both models, therefore, describe merely the asymptotic cases of interfacial phonon behaviour.

The temperature jump expression is given by

$$\Delta T = R_k \frac{\dot{Q}_{1\to 2}(T_2) - \dot{Q}_{1\to 2}(T_1)}{A} \qquad (27)$$

where $R_k$ refers to the Kapitza resistance, $A$ is the interfacial area, and $\dot{Q}_{1\to 2}(T)$ is the heat current between the two media, which is assumed here to be independent of the temperature on the other side of the interface to simplify the analysis. In light of the thermal rectification effect observed in MD simulations, this assumption may be invalid.

The heat current comprises the total phonon energy being transmitted across the interface and can be evaluated from the expression

$$\frac{\dot{Q}_{1\to 2}}{A} = \frac{1}{2} \sum_j \int_0^{\omega_1^{max}} \int_0^{\pi/2} N_{1,j} \hbar \omega c_{1,j} \alpha_{1\to 2} \cos\theta \sin\theta \, d\theta \, d\omega \qquad (28)$$

where $N_{1,j}$ is the density of phonon states, $\hbar\omega$ is the phonon energy, $c_{1,j}$ is the phonon velocity with subscript $i$ indicating the medium, $j$ is the phonon mode, $\alpha_{1\to 2}$ is the transmission probability, and $\theta$ is the incident angle.

The AMM and DMM models differ only in their respective forms of the transmission probability. In the AMM model, the transmission probability for a normal incident angle can be obtained from continuum acoustic theory as

$$\alpha_{1\to 2} = \frac{4 Z_1 Z_2}{(Z_1 + Z_2)^2} \qquad (29)$$

where $Z_1$ and $Z_2$ denote the respective acoustic impedance of each medium.

The transmission probability used in the DMM model is based on a Debye approximation for the phonon velocities and density of states

$$\alpha_i(\omega) = \frac{\sum_j c_{3-i,j}^{-2}}{\sum_j c_{i,j}^{-2}} . \qquad (30)$$

The predictions of the thermal boundary resistance given by the aforementioned models as well as other improved models such as the scattering-mediated AMM by Prasher & Phelan [156] deviated rather significantly from experimentally observed values [138]. The poor agreement may be attributed to the neglected influence of interfacial molecular parameters, the breakdown of the Debye approximation at high temperatures, and the assumption that heat transfer in liquids can be adequately described by phonon theory [157].

## 13 Measurement of Gas Temperature Jump

The earliest experimental verification of the temperature jump phenomena was performed by Smoluchowski [149] through the measurements of heat conduction between two parallel surfaces at different temperatures for air and hydrogen. He observed that the temperature jump distance was proportional to pressure, or equivalently the mean free path. His findings were later corroborated by other researchers using nearly similar methods [158].

Other temperature jump or accommodation coefficient measurement techniques include the popular hot-wire method, which measures the amount of energy required to maintain an electrically heated wire immersed in the test gas at a given temperature and determination of thermal conductivity of powder beds in gases [70]. These early experimental investigations have been reviewed comprehensively in the literature [158]. More recently, Trott *et al*. [159] employed an updated parallel-plate setup that was housed in a vacuum chamber for two different accommodation coefficient measurement approaches. The first method was to obtain the heat-flux indirectly through temperature difference while the second involved the measurement of gas density profiles by electron-beam fluorescence, which can then be converted to temperature profiles. High-precision instruments were installed to control factors including gas pressure, gas and plate temperature, and fluorescence detection.

## 14 Interfacial Thermal Resistance (Temperature Jump)

Experimental work on the temperature discontinuity or the equivalent thermal boundary resistance took off in the mid-20th century after it was proposed that a thermal resistance could exist between liquid helium and a solid surface. Incipient studies on the thermal boundary resistance are chronicled in two review papers [137,138]. The first reported measurement of temperature drop at a liquid-solid interface was performed by Kapitza [10] (hence eponymously termed Kapitza resistance) using a simple technique of measuring the temperature profile around the interfacial region between a copper specimen and liquid helium at temperatures below $1K$. As helium is in a superfluid state with negligible thermal conductivity at such temperatures, its temperature could be taken from any location within the liquid while the temperature profile within the copper was extrapolated up to the interface using several thermometers. This bypassed the difficulty of probing the temperatures at both sides of the interface. Later, an indirect approach was developed to evaluate the thermal resistance from the amplitudes of the transmitted and reflected second-sound wave that is incident on a thin metal foil that was immersed within liquid helium. The propagation of heat in superfluid helium occurs through the second-sound. The detection of a reflected sound wave at the interface indicates a finite thermal resistance.

To the best of our knowledge, only one active research group has been conducting experiments on the thermal conductance (inverse of thermal resistance) of liquid-solid interfaces at room temperature. In their original experiment, the thermal conductance was obtained from the cooling curves of metallic nanoparticle suspensions, which were measured using pump-probe laser spectroscopy [160]. This method was later realised to be inappropriate for investigating the effect of wetting due to the clustering of hydrophobic particles. Subsequently, time-domain thermoreflectance was employed to study the thermal conductance of planar interfaces between water and functionalised metal substrates through the fitting of an analytical heat transfer model to reflectivity curves [143]. The drawbacks of this technique include the high experimental uncertainty due to the inaccurate determination of film thickness and heat capacity, as well as additional thermal resistances which could arise from substrate contamination and electron-phonon coupling.

Recent studies on thermal resistance comprise of MD simulations, the bulk of which focus on the role of wetting and surface roughness [142,144,145,147,161]. Apart from the general shortcomings of the MD method listed previously, another criticism lies in its classical nature, thereby not only limiting the accuracy at low temperatures but also the inability to consider the influence of electrons [139].

## 15  Summary and Views

In this paper, we described the mechanisms that are thought to be the cause of the fluid-solid boundary jump of velocity and temperature. Based on these proposed mechanisms, several theoretical models have been developed but are mostly inadequate in providing the accurate predictions of experimentally observed trends. The series of experimental techniques that have been reviewed here show great novelty in overcoming the difficulty of indirect measurements. However, results from these high-resolution methods have to be interpreted with caution as they often contain the inherent sources of apparent effects, consequently presenting a misleading picture of the interfacial phenomena. On the other hand, such unintended effects may be useful as a form of artificial control of the jump behaviour in small-scale devices. The MD simulations of simple flow and heat transfer systems allow the study of the relation between molecular behaviour and the macroscopic discontinuity across the interface but outcomes are highly dependent on prescribed input conditions. Besides, simulated variables do not translate to realistic values for practical comparisons.

The confounding information gathered from experiments and simulations deserves to be addressed theoretically in greater detail. In addition, the largely similar characteristics of the respective boundary conditions beg the question of whether the interfacial jump phenomenon originates from a common physical mechanism. If so, this would indicate that a single general boundary condition model should apply to both gases and liquids [16,17]. A critical issue herein is whether observed interfacial behaviour arises from molecular interactions, secondary processes, or more likely, a combination of the two.

**Acknowledgement**

This work was supported by Nanyang Technological University (M4081942).

Source: Applied Mechanics Reviews, Vol. 69, No. 2, pp. 020801, 2017;
DOI: 10.1115/1.4036191[24] Gomes, K.K., Mar, W., Ko, W., Guinea, F., and Manoharan, H.C., 2012, "Designer Dirac Fermions and Topological Phases in Molecular Graphene," *Nature*, **483**(7389), pp. 306–310.

[25] Israelachvili, J.N., 2012, *Intermolecular and Surface Forces*, 3rd edition, World Publishing Corporation Beijing Company.

[26] Cheng, L., Fenter, P., Nagy, K.L., Schlegel, M.L., and Sturchio, N.C., 2001, "Molecular-Scale Density Oscillations in Water Adjacent to a Mica Surface," *Physical Review Letters*, **87**(15), pp. 156103.

[27] Barrat, J.-L., and Bocquet, L., 1999, "Influence of Wetting Properties on Hydrodynamic Boundary Conditions at a Fluid/Solid Interface," *Faraday Discussions*, **112**, pp. 119–127.

[28] Groß, A., 2009, *Theoretical Surface Science: A Microscopic Perspective*, Second Edition, Springer.

[29] Brenner, H., and Ganesan, V., 2000, "Molecular Wall Effects: Are Conditions at a Boundary "Boundary Conditions"?" *Physical Review E*, **61**(6B), pp. 6879–6897.

[30] Cucchetti, A., and Ying, S.C., 1996, "Memory Effects in the Frictional Damping of Diffusive and Vibrational Motion of Adatoms," *Physical Review B*, **54**(5), pp. 3300–3310.

[31] de Gennes, P.G., 2002, "On Fluid/Wall Slippage," *Langmuir*, **18**(9), pp. 3413–3414.

[32] Gutfreund, P., Wolff, M., Maccarini, M., Gerth, S., Ankner, J.F., Browning, J., Halbert, C.E., Wacklin, H., and Zabel, H., 2011, "Depletion at Solid/Liquid Interfaces: Flowing Hexadecane on Functionalized Surfaces," *Journal of Chemical Physics*, **134**(6), pp. 064711.

[33] Ala-Nissila, T. Ferrando, R., and Ying, S.C., 2002, "Collective and Single Particle Diffusion on Surfaces," *Advances in Physics*, **51**(3), pp. 949–1078.

[34] Richardson, S., 1973, "On the No-Slip Boundary Condition," *Journal of Fluid Mechanics*, **59**(4), pp. 707–719.

[35] Dussan, E.B., and Davis, S.H., 1974, "On the Motion of a Fluid-Fluid Interface along a Solid Surface," *Journal of Fluid Mechanics*, **65**(1), pp. 71–95.

[36] Brigo, L., Natali, M., Pierno, M., Mammano, F., Sada, C., Fois, G., Pozzato, A., dal Zilio, S., Tormen, M., and Mistura, G., 2008, "Water Slip and Friction at a Solid Surface," *Journal of Physics-Condensed Matter*, **20**(35), pp. 354016.

[37] Cottin-Bizonne, C., Barrat, J.-L., Bocquet, L., and Charlaix, E., 2003, "Low-Friction Flows of Liquid at Nanopatterned Interfaces," *Nature Materials*, **2**(4), pp. 237–240.

[38] Bonaccurso, E., Butt, H.-J., and Craig, V.S.J., 2003, "Surface Roughness and Hydrodynamic Boundary Slip of a Newtonian Fluid in a Completely Wetting System," *Physical Review Letters*, **90**(14), pp. 144501.

[39] Truesdell, R., Mammoli, A., Vorobieff, P., van Swol, F., and Brinker, C.J., 2006, "Drag Reduction on a Patterned Superhydrophobic Surface," *Physical Review Letters*, **97**(4), pp. 044504.

[40] Sbragaglia, M., Benzi, R., Biferale, L., Succi, S., and Toschi, F., 2006, "Surface Roughness-Hydrophobicity Coupling in Microchannel and Nanochannel Flows," *Physical Review Letters*, **97**(20), pp. 204503.25

Source: Applied Mechanics Reviews, Vol. 69, No. 2, pp. 020801, 2017;
DOI: 10.1115/1.4036191[113]  Honig, C.D.F., and Ducker, W.A., 2007, "No-Slip Hydrodynamic Boundary Condition for Hydrophilic Particles," *Physical Review Letters*, **98**(2), pp. 028305.

[114]  Henry, C.L., and Craig, V.S.J., 2009, "Measurement of No-Slip and Slip Boundary Conditions in Confined Newtonian Fluids Using Atomic Force Microscopy," *Physical Chemistry Chemical Physics*, **11**(41), pp. 9514–9521.

[115]  Lauga, E., 2004, "Apparent Slip due to the Motion of Suspended Particles in Flows of Electrolyte Solutions," *Langmuir*, **20**(20), pp. 8924–8930.

[116]  Li, Z., D'eramo, L., Monti, F., Vayssade, A.-L., Chollet, B., Bresson, B., Tran, Y., Cloitre, M., and Tabeling, P., 2014, "Slip Length Measurements Using mu PIV and TIRF-Based Velocimetry," *Israel Journal of Chemistry*, **54**(11-12), pp. 1589–1601.

[117]  Joly, L., Ybert, C., and Bocquet, L., 2006, "Probing the Nanohydrodynamics at Liquid-Solid Interfaces Using Thermal Motion," *Physical Review Letters*, **96**(4), pp. 046101.

[118]  Daikhin, L., Gileadi, E., Tsionsky, V., Urbakh, M., and Zilberman, G., 2000, "Slippage at Adsorbate-Electrolyte Interface. Response of Electrochemical Quartz Crystal Microbalance to Adsorption," *Electrochimica Acta*, **45**(22-23), pp. 3615–3621.

[119]  Du, B., Goubaidoulline, I., and Johannsmann, D., 2004, "Effects of Laterally Heterogeneous Slip on the Resonance Properties of Quartz Crystals Immersed in Liquids," *Langmuir*, **20**(24), pp. 10617–10624.

[120]  McHale, G., and Newton, M.I., 2004, "Surface Roughness and Interfacial Slip Boundary Condition for Quartz Crystal Microbalances," *Journal of Applied Physics*, **95**(1), pp. 373–380.

[121]  Willmott, G.R., and Tallon, J.L., 2007, "Measurement of Newtonian Fluid Slip Using a Torsional Ultrasonic Oscillator," *Physical Review E*, **76**(6), pp. 066306.

[122]  Churaev, N.V., Ralston, J., Sergeeva, I.P., and Sobolev, V.D., 2002, "Electrokinetic Properties of Methylated Quartz Capillaries," *Advances in Colloid and Interface Science*, **96**(1-3), pp. 265–278.

[123]  Watanabe, K., Takayama, T., Ogata, S., and Isozaki, S., 2003, "Flow between two Coaxial Rotating Cylinders with a Highly Water-Repellent Wall," *AIChE Journal*, **49**(8), pp. 1956–1963.

[124]  Perisanu, S., and Vermeulen, G., 2006, "Curvature, Slip, and Viscosity in He-3-He-4 Mixtures," *Physical Review B*, **73**(13), pp. 134517.

[125]  Bocquet, L., Tabeling, P., and Manneville, S., 2006, "Comment on "Large Slip of Aqueous Liquid Flow over a Nanoengineered Superhydrophobic Surface"," *Physical Review Letters*, **97**(10), pp. 109601.

[126]  Harting, J., Kunert, C., and Hyvaluoma, J., 2010, "Lattice Boltzmann Simulations in Microfluidics: Probing the No-Slip Boundary Condition in Hydrophobic, Rough, and Surface Nanobubble Laden Microchannels," *Microfluidics and Nanofluidics*, **8**(1), pp. 1–10.

[127]  Pahlavan, A.A., and Freund, J.B., 2011, "Effect of Solid Properties on Slip at a Fluid-Solid Interface," *Physical Review E*, **83**(2), pp. 021602.

[128]  Yong, X., and Zhang, L.T., 2013, "Slip in Nanoscale Shear Flow: Mechanisms of Interfacial Friction," *Microfluidics and Nanofluidics*, **14**(1-2), pp. 299–308.

[129]  Arkilic, E.B., Schmidt, M.A., and Breuer, K.S., 1997, "Gaseous Slip Flow in Long Microchannels," *Journal of Microelectromechanical Systems*, **6**(2), pp. 167–178.
30

Source: Applied Mechanics Reviews, Vol. 69, No. 2, pp. 020801, 2017;
DOI: 10.1115/1.4036191[130] Harley, J.C., Huang, Y., Bau, H.H., and Zemel, J.N., 1995, "Gas Flow in Micro-Channels," *Journal of Fluid Mechanics*, **284**, pp. 257–274.

[131] Bentz, J.A., Tompson, R.V., and Loyalka, S.K., 2001, "Measurements of Viscosity, Velocity Slip Coefficients, and Tangential Momentum Accommodation Coefficients Using a Modified Spinning Rotor Gauge," *Journal of Vacuum Science & Technology A-Vacuum Surfaces and Films*, **19**(1), pp. 317–324.

[132] Bentz, J.A., Tompson, R.V., and Loyalka, S.K., 1999, "Viscosity and Velocity Slip Coefficients for Gas Mixtures: Measurements with a Spinning Rotor Gauge," *Journal of Vacuum Science & Technology A-Vacuum Surfaces and Films*, **17**(1), pp. 235–241.

[133] Graur, I.A., Perrier, P., Ghozlani, W., and Méolans, J.G., 2009, "Measurements of Tangential Momentum Accommodation Coefficient for Various Gases in Plane Microchannel," *Physics of Fluids*, **21**(10), pp. 102004.

[134] Bird, G.A., 1994, *Molecular Gas Dynamics and the Direct Simulation of Gas Flows*, 2nd edition, Clarendon Press.

[135] Shen, C., 2010, *Rarefied Gas Dynamics: Fundamentals, Simulations and Micro Flows*, Springer.

[136] Kennard, E.H., 1954, *Kinetic Theory of Gases*, McGraw-Hill.

[137] Pollack, G.L., 1969, "Kapitza Resistance," *Reviews of Modern Physics*, **41**(1), pp. 48–81.

[138] Swartz, E.T., and Pohl, R.O., 1989, "Thermal Boundary Resistance," *Reviews of Modern Physics*, **61**(3), pp. 605–668.

[139] Cahill, D.G., Ford, W.K., Goodson, K.E., Mahan, G.D., Majumdar, A., Maris, H.J., Merlin, R., and Phillpot, S.R., 2003, "Nanoscale Thermal Transport," *Journal of Applied Physics*, **93**(2), pp. 793–818.

[140] Goicochea, J.V., Hu, M., Michel, B., and Poulikakos, D., 2011, "Surface Functionalization Mechanisms of Enhancing Heat Transfer at Solid-Liquid Interfaces," *Journal of Heat Transfer-Transactions of the ASME*, **133**(8), pp. 082401.

[141] Acharya, H., Mozdzierz, N.J., Keblinski, P., and Garde, S., 2012, "How Chemistry, Nanoscale Roughness, and the Direction of Heat Flow Affect Thermal Conductance of Solid-Water Interfaces," *Industrial & Engineering Chemistry Research*, **51**(4), pp. 1767–1773.

[142] Wang, Y., and Keblinski, P., 2011, "Role of Wetting and Nanoscale Roughness on Thermal Conductance at Liquid-Solid Interface," *Applied Physics Letters*, **99**(7), pp. 073112.

[143] Ge, Z., Cahill, D.G., and Braun, P.V., 2006, "Thermal Conductance of Hydrophilic and Hydrophobic Interfaces," *Physical Review Letters*, **96**(18), pp. 186101.

[144] Shenogina, N., Godawat, R., Keblinski, P., and Garde, S., 2009, "How Wetting and Adhesion Affect Thermal Conductance of a Range of Hydrophobic to Hydrophilic Aqueous Interfaces," *Physical Review Letters*, **102**(15), pp. 156101.

[145] Murad, S., and Puri, I.K., 2008, "Thermal Transport across Nanoscale Solid-Fluid Interfaces," *Applied Physics Letters*, **92**(13), pp. 133105.

[146] Xue, L., Keblinski, P., Phillpot, S.R., Choi, S.U.-S., and Eastman, J.A., 2003, "Two Regimes of Thermal Resistance at a Liquid-Solid Interface," *Journal of Chemical Physics*, **118**(1), pp. 337–339.
31

Source: Applied Mechanics Reviews, Vol. 69, No. 2, pp. 020801, 2017;
DOI: 10.1115/1.4036191[147] Hu, M., Goicochea, J.V., Michel, B., and Poulikakos, D., 2009, "Thermal Rectification at Water/Functionalized Silica Interfaces," *Applied Physics Letters*, **95**(15), pp. 151903.

[148] Murad, S., and Puri, I.K., 2012, "Communication: Thermal Rectification in Liquids by Manipulating the Solid-Liquid Interface," *Journal of Chemical Physics*, **137**(8), pp. 081101.

[149] Smoluchowski, M.S., 1898, "Ueber Wärmeleitung in Verdünnten Gasen," *Annalen der Physik*, **300**(1), pp. 101–130.

[150] Schäfer, K., Rating, W., and Eucken, A., 1942, "Influence of the Inhibited Exchanges of Translation and Vibration Energy to Heat Conduction of Gases," *Annalen der Physik*, **42**(2/3), pp. 176–202.

[151] Dadzie, S.K., and Méolans, J.G., 2005, "Temperature Jump and Slip Velocity Calculations from an Anisotropic Scattering Kernel," *Physica A-Statistical Mechanics and its Applications*, **358**(2-4), pp. 328–346.

[152] Baule, B., 1914, "Theoretische Behandlung der Erscheinungen in Verdünnten Gasen," *Annalen der Physik*, **44**(1), pp. 145–176.

[153] Deissler, R.G., 1964, "An Analysis of Second-Order Slip Flow and Temperature-Jump Boundary Conditions for Rarefied Gases," *International Journal of Heat and Mass Transfer*, **7**(6), pp. 681–694.

[154] Lees, L., and Liu, C.-Y., 1960, *Kinetic Theory Description of Plane, Compressible Couette Flow*, California Institute of Technology.

[155] Mazo, R.M., 1955, *Theoretical Studies on Low Temperature Phenomena*, Yale University.

[156] Prasher, R.S., and Phelan, P.E., 2001, "A Scattering-Mediated Acoustic Mismatch Model for the Prediction of Thermal Boundary Resistance," *Journal of Heat Transfer-Transactions of the ASME*, **123**(1), pp. 105–112.

[157] Bolmatov, D., Brazhkin, V.V., and Trachenko, K., 2012, "The Phonon Theory of Liquid Thermodynamics," *Scientific Reports*, **2**, pp. 421.

[158] Devienne, F.M., 1965, "Low Density Heat Transfer," *Advances in Heat Transfer*, **2**, pp. 271–356.

[159] Trott, W.M., Rader, D.J., Castañeda, J.N., Torczynski, J.R., and Gallis, M.A., 2008, "Measurement of Gas-Surface Accommodation," *Proceedings of the 26th International Symposium*, pp. 621–628.

[160] Ge, Z., Cahill, D.G., and Braun, P.V., 2004, "AuPd Metal Nanoparticles as Probes of Nanoscale Thermal Transport in Aqueous Solution," *Journal of Physical Chemistry B*, **108**(49), pp. 18870–18875.

[161] Kim, B.H., Beskok, A., and Cagin, T., 2008, "Molecular Dynamics Simulations of Thermal Resistance at the Liquid-Solid Interface," *Journal of Chemical Physics*, **129**(17), pp. 174701.
32